\documentclass[a4paper,UKenglish]{lipics}
 
\usepackage{microtype}

\title{Transitioning to Proof}
\titlerunning{Transitioning to Proof} 

\author[1]{Diane Resek}
\author[2]{Dan Fendel}
\affil[1]{Department of Mathematics, San Francisco State University\\
1600 Holloway Ave, San Francisco, USA\\
  \texttt{resek@sfsu.edu}}
\affil[2]{Department of Mathematics, San Francisco State University\\
  1600 Holloway Ave, San Francisco, USA\\
  \texttt{dfendel@sfsu.edu}}
\authorrunning{D. Resek and D. Fendel} 

\Copyright{Diane Resek and Dan Fendel}


\serieslogo{logo_ttl}
\volumeinfo
  {M. Antonia {Huertas}, Jo\~ao {Marcos}, Mar\'ia {Manzano}, Sophie {Pinchinat}, \\
  Fran\c{c}ois {Schwarzentruber}}
  {5}
  {4th International Conference on Tools for Teaching Logic}
  {1}
  {1}
  {185}
\EventShortName{TTL2015}

\begin{document}

\maketitle

\begin{abstract}
This paper describes some strategies used in a ‘transition’ course. Such courses help undergraduate mathematics majors move from learning procedures to learning to function as  critical mathematicians in order to understand and work with abstract concepts.   One of the co-authors of this paper was a student of Leon Henkin. His influence on her helped shape the strategies used in the course, and is described at the end of the paper. 
 \end{abstract}

\section{Introduction}
In the United States we often hear students complain: “I understand the math, but I just can’t write proofs.” A few probing questions usually reveal that the student really does not understand the mathematics involved. She or he may be able to parrot back the key definitions, but doesn’t know what inferences can be drawn or what relationship needs to be proved. What is really happening here?
First, students often believe that they understand mathematical concepts after they have read about them in the text or attended a lecture on the material. They believe that, if they find they can’t do the homework problems, whether proofs or other non-mechanical exercises, then there must be something wrong with the book or the lecture.
Students need to learn what people who have become mathematicians instinctively know and probably can’t remember being taught. That is, in order to understand a new idea, they must play with it, examine examples, look for counterexamples, ask themselves questions about it, and more. 
Currently, many universities in the United States offer courses to students before they take proof oriented courses such as abstract algebra or real analysis. Such courses are designed to help students to “transition” between courses that focus on procedures, such as calculus as taught in the U.S., and more conceptually oriented courses. The courses are aptly named “transition courses.”
The authors of this paper wrote a text \cite{Fendel-Resek} for such a course, 1990). The text and the course sought to help students learn how to cope with open-ended, unstructured questions; how to formulate conjectures; how to evaluate the reasonableness of a statement; how to make plausibility arguments; how to make constructive use of examples; etc. These skills are difficult to learn, and some time needs to be spent acquiring them when the student and the instructor do not have other important agendas.
One of the authors of this paper and the text, Diane Resek, was a PhD student of Leon Henkin, and worked closely with him on mathematics education projects as well as mathematical logic. This work shaped her thinking about logic and the teaching of logic. She has written about some of her intellectual inheritance from Henkin in the chapter “Lessons from Leon” in the book, The Life and Work of Leon Henkin \cite{Manzano}.
In this paper we will discuss some of the techniques used in the transition course and its text and then explain how the influence of Leon Henkin helped shape the ideas behind them. The techniques are:
\begin{itemize}
\item emphasizing exploration;
\item including exercises entitled “Get your hands dirty”;
\item introducing logic connectors via the idea of counter-examples and not truth tables; and
\item asking students to judge the correctness of proofs.
\end{itemize}

We will discuss each of these elements in turn.

\section{Exploration}

One way to motivate students’ interest in proof writing is to raise questions as to the truth of a statement. This method is especially motivating if students are engaged in arguments with their peers about the correctness of the statement. The course begins with the presentation of the rules for a simple game. Students play the game in groups and are asked to raise questions about the game. They come up with a number of questions, but settle on trying to find a winning strategy. The activity ends with students trying to prove that a given strategy works. For homework they work on strategies for other games.

As the course progresses, they meet new subject matter such as sets, integers, quantifiers, functions and sequences. Each time a new subject is introduced, students work at exploring that subject before they begin writing proofs in that domain. For example, when quantifiers and connectives are introduced, students are asked:

Consider “($\exists x \in S)(p(x) $and $q(x))$" . How does that statement compare to “$(\exists x \in S)(p(x)) $and$ (\exists x\in S)(q(x)))$”? Can the original statement be true and the combined statement false? Or vice versa? Try some examples with specific open sentences and look for general principles.

Students’ feelings of frustration are acknowledged in the text. They are not used to being directed away from collecting a set of answers and toward the process of thinking about mathematics. They are not used to being left with many open problems in a mathematics course. We let them know that, for this course and for their future work with mathematics, they need to get used to the fact that they will sometimes be coming away from their work with more questions than answers. We remind them that they can always return to their questions on another day, but we emphasize that coming away with good questions is a positive accomplishment.

The object of exploration in our text and course is not simply to motivate students to prove things, but rather, to build their intuition about a given topic. We seek to instill in students a habit of exploring on their own each time they come up against a new mathematical concept.

\section{Get Your Hands Dirty}

Throughout the text, students meet exercises labeled: “Get Your Hands Dirty” (GYHD). They are told “This means you must stop being an armchair mathematician and get to work.” They are also told:

Mathematics is not a spectator sport. Even when you are just reading mathematics, you need to be an active participant, or you won’t understand. Advanced mathematics texts expect you to explore, though they don’t explicitly tell you to do this. For example, when a mathematics book gives you a new definition, you must “play with it.” That’s the best way you can come to understand a new idea. Advanced textbooks will expect you to have an understanding of the new idea as they proceed.

We don’t expect students to know how to play with definitions when they come to our transition course. We try to teach them that. For instance, when we introduce the definition of a power set, we give them examples. Then we give them a GYHD exercise in which they are asked to find the power sets of some small sets, and later ask them (in another GYHD) whether something can be simultaneously both an element of and a subset of some other set. Similarly, when quantifiers are introduced, students are given the definition of what it means to say that set A is a subset of set $B$: $(\forall x\in A)(x\in B)$. They are then asked, in a GYHD, to come up with an expression using quantifiers for what it means to say that A is not a subset of $B$. 

GYHDs are different from exploration exercises in that GYHDs are generally very short activities through which students come to understand a theorem or a definition. The object of GYHDs is to train students to “play” with new definitions and theorems. For instance, they need to learn to ask themselves why certain qualifications such as “$x \neq 0$” are given, or why the definition is not more general. An “exploration” is a longer exercise in which students are expected to make conjectures and then try to prove or disprove them.

Since we used GYHDs when introducing new definitions, we expected that students would learn how to make sense of new definitions by the end of the course. So, on the final examination, we gave students the following formal definition:
A set of real numbers $S$ is called connected if 
$\forall x\forall y\forall z(x\in S$ and $y\in S$ and $x\leq z\leq y \rightarrow z\in S)$

We expected the definition to be totally new for them, since the concept doesn’t come up in earlier courses. We intentionally did not give them any context for the definition, because we wanted them to decipher it on their own.
We then asked them to come up with examples of sets of real numbers that fit the definition and sets that did not. The goal was for them to learn how to build an intuitive understanding based on a formal definition. Finally, we asked them to prove a specific statement that followed easily from the definition.

\section{Logic Without Truth Tables}

Probably everyone teaching the truth table of the conditional connective has come up against resistance from students to the notion that A→B is considered true anytime A is false. This idea simply does not make sense to many students. We therefore decided to ignore truth tables in the main part of the book, and work instead with predicate logic. That is, instead of working with truth tables, we emphasized the role of counterexamples in assessing the truth of conditional sentences. So, for our purposes, the key idea is that a conditional sentence is true unless there is a counterexample. Proofs of conditional statements then become demonstrations that there is no counterexample. 

Of course, as one would hope, this definition of the truth of conditionals is equivalent to the one given by truth tables. But in this approach, we did not meet the usual resistance to the definition or the use of the definition.

Thus, students see that to prove that a conditional statement is true, one can assume that we have some unspecified object that makes the hypothesis true, and we then need to show that the conclusion must be true. If we can do this (knowing nothing about the object except that it makes the hypothesis true), then we have shown that there can be no counterexample. 

They can also use and justify proof by contrapositive in the same way: start with an unspecified object that makes the conclusion false and demonstrate that the object makes the hypothesis false, and thus, again, the object is not a counterexample. Similarly, proof by contradiction is justified by assuming one has a counterexample and showing that this leads to an impossible situation. Thus, there can be no counterexample.
Note: We did put some more formal logic in an appendix to make the book useful to a wider audience of instructors.

\section{Proof Evaluation}

We found that a particularly useful exercise in helping students develop their proof writing abilities was to give them “potential proofs” and ask them to evaluate whether the given piece of writing was an actual proof. This meant that they needed first to decide whether the “theorem” (i.e., what was supposedly being proved) was in fact true, and, if so, to then determine whether the given “proof” was in fact valid. If it was not valid, they were to identify the flaw. 

Here’s an early example: [“$P(A)$” is our notation for the power set of $A$.]

“Theorem”: For any sets $A$ and $B$, if $P(A)\cap P(B) \neq \emptyset$ then $A \cap B \neq \emptyset$.

“Proof” Assume $P(A)\cap P(B) \neq \emptyset$. Our goal is to show $A \cap B \neq \emptyset$. Since $P(A) \cap P(B)$ is non-empty, there is something in this intersection; call it $X$. So, $X \subseteq A$ and $X \subseteq B$. Now let y be any element of $X$. So $y \in A \cap B$, and so $A \cap B \neq \emptyset$. This concludes the proof.

This type of exercise helped students to understand some subtleties of proof and alerted them to some pitfalls that they might come across. We wanted students to distinguish between “omissions of detail” and real mistakes. (For instance, in this “proof,” we could have stated the definition of power set in going from saying that $X$ is in $P(A)\cap P(B)$ to saying $X \subseteq A$ and $X \subseteq B$. That is simply an “omission of detail” but the step is a valid one) .

Often it was helpful to ask students to look for a counterexample, and then find the first statement in the “proof” that would actually be false for that counterexample.

\section{The Influence of Leon Henkin: Personal Reflections from Diane Resek}

Each of the techniques discussed here—exploration, “Get Your Hands Dirty,” use of predicate logic rather than propositional logic, and “proof evaluation”—was designed to make students’ work with mathematical ideas more intuitive, more natural, and more like what mathematicians actually do. 

Leon always sought in his own teaching to make ideas intuitive. As a thesis advisor he pushed me to try new approaches and to look at concrete examples whenever possible. In my dissertation work on infinite dimensional cylindric algebras, his approach led me to think of each dimension as a separate piece of paper. I actually wrote out some examples on a large number of pages and this helped me come to a major conjecture. 

Those pieces of paper worked for me. This experience and others—seeing the power of getting one’s hands dirty while working with concrete examples—led me to use the strategy while teaching students. 

Another influence of Henkin on me, and thus on the transition course, came from his drive to make mathematical knowledge available to as wide an audience as possible. He said that although there might be a level of mathematical knowledge that average people could not attain, he believed that this level was beyond what is needed to complete an undergraduate mathematics major. One goal of the transition course was to make the higher-level courses in the mathematics major accessible to more students.





\newpage
\thispagestyle{empty}
{\ }

\end{document}